\documentstyle[emulateapj,apjfonts,psfig]{article}

\def\grb{GRB\thinspace{000301C}}

\begin{document}
 
\title{\large \bf A JET MODEL FOR THE AFTERGLOW EMISSION FROM GRB\,000301C}

\author{ 
 E. Berger\altaffilmark{1},
 R. Sari\altaffilmark{2},
 D. A. Frail\altaffilmark{3}, 
 S. R. Kulkarni\altaffilmark{1},
 F. Bertoldi\altaffilmark{4},
 A. Peck\altaffilmark{4}, 
 K. Menten\altaffilmark{4},
 D. S. Shepherd\altaffilmark{3},
 G. H. Moriarty-Schieven\altaffilmark{5},
 G. Pooley\altaffilmark{6},
 J. S. Bloom\altaffilmark{1},
 A. Diercks\altaffilmark{1},
 T. J. Galama\altaffilmark{1},
 K. Hurley\altaffilmark{7}}

\altaffiltext{1}{California Institute of Technology, 
Owens Valley Radio Observatory
105-24, Pasadena, CA 91125}

\altaffiltext{2}{California Institute of Technology,
 Theoretical Astrophysics  130-33, Pasadena, CA 91125}          

\altaffiltext{3}{National Radio Astronomy Observatory, P.~O.~Box O,
  Socorro, NM 87801}

\altaffiltext{4}{Max-Planck-Institut fuer Radioastronomie,
 Auf dem Huegel 69, D-53121 Bonn}

\altaffiltext{5}{Joint Astronomy Centre, 660 N. A'ohoku Place Hilo, HI
96720}

\altaffiltext{6}{Mullard Radio Astronomy Observatory, Cavendish
  Laboratory, Madingley Road, Cambridge CB3 0HE}      

\altaffiltext{7}{University of California, Berkeley, Space Sciences
  Laboratory, Berkeley, CA 94720-7450}

\begin{abstract}
  We present broad-band radio observations of the afterglow of \grb{},
  spanning from 1.4 to 350 GHz for the period of 3 to 83 days after
  the burst.  This radio data, in addition to measurements at the
  optical bands, suggest that the afterglow arises from a collimated
  outflow, i.e. a jet. To test this hypothesis in a self-consistent
  manner, we employ a global fit and find that a model of a jet,
  expanding into a constant density medium (ISM+jet), provides the
  best fit to the data. A model of the burst occurring in a wind-shaped
  circumburst medium (wind-only model) can be ruled out, and a
  wind+jet model provides a much poorer
  fit of the optical/IR data than the ISM+jet model.  In addition, we
  present the first clear indication that the reported fluctuations in
  the optical/IR are achromatic with similar amplitudes in all bands, and 
  possibly extend into the radio regime.  Using the parameters derived from
  the global fit, in particular a jet break time, $t_{jet} \approx 7.5$
  days, we infer a jet opening angle of $\theta_0 \approx 0.2$, and
  consequently the estimate of the emitted energy in the GRB itself is
  reduced by a factor $\sim$50 relative to the isotropic value, giving
  $E\approx 1.1 \times 10^{51}$ ergs.
\end{abstract}

\keywords{gamma rays:bursts -- radio continuum:general --
  cosmology:observations}

\section{Introduction}\label{sec:intro}

\grb\ is the latest afterglow to exhibit a break in its optical/IR
light curves. An achromatic steepening of the light curves has been
interpreted in previous events (e.g. \cite{kdo+99}; \cite{H+99}) as
the signature of a jet-like outflow (\cite{rho99}; \cite{SPH99}),
produced when relativistic beaming no longer ``hides'' the
non-spherical surface, and when the ejecta undergo rapid lateral
expansion. The question of whether the relativistic outflows from
gamma-ray bursts (GRBs) emerge isotropically or are collimated in jets
is an important one. The answer has an impact both on estimates of the
GRB event rate and the total emitted energy --- issues that have a
direct bearing on GRB progenitor models.

An attempt by Rhoads and Fruchter (2000) to model this break using
only the early time ($\Delta t \lesssim 14$ days) optical/IR data has
led to a jet interpretation of the afterglow evolution, but with
certain peculiar aspects, such as a different jet break time at R band
than at K' band. However, subsequent papers by Masetti et al.~(2000),
and Sagar et al. (2000), with larger optical data sets, pointed out
that there are large flux density variations ($\sim 30\%$) on
timescales as short as a few hours, superposed on the overall
steepening of the optical/IR light curves.  While the origin of these
peculiar fluctuations remains unknown, it is clear that they
complicate the fitting of the optical/IR data, rendering some of the
Rhoads and Fruchter results questionable.

In this paper we take a different approach. We begin by presenting
radio measurements of this burst from 1.4 GHz to 350 GHz, spanning a
time range from 3 to 83 days after the burst.  These radio
measurements, together with the published optical/IR data, present a
much more comprehensive data set, which is less susceptible to the
effects of the short-timescale optical fluctuations.  We then use the
entire data set to fit a global, self-consistent jet model, and derive
certain parameters of the GRB from this model.
Finally, we explore the possibility of a wind, and wind+jet global fit
to the data, and compare our results with the conclusions drawn in the
previous papers.

\section{Observations}
\label{sec:obs}

Radio observations were made from 1.43 GHz to 350 GHz, at a number of
facilities, including the James Clark Maxwell Telescope 
(JCMT\footnotemark\footnotetext{The JCMT is operated by The Joint
  Astronomy Centre on behalf of the Particle Physics and Astronomy
  Research Council of the UK, the Netherlands Organization for
  Scientific Research, and the National Research Council of Canada.}),
the Institut f$\ddot{\rm u}$r RadioAstronomie im Millimeterbereich (IRAM
\footnotemark\footnotetext{IRAM is supported by INSU/CNRS (France), 
MPG (Germany) and IGN (Spain).}),
the Owens Valley Radio Observatory Interferometer (OVRO), the Ryle
Telescope and the (Very Large Array (VLA\footnotemark\footnotetext{The
    NRAO is a facility of the National Science Foundation operated
    under cooperative agreement by Associated Universities, Inc.  NRAO
    operates the VLA.}). A log of these observations and the flux
  density measurements are summarized in Table~\ref{tab:Table-VLA}.
  With the exception of IRAM, we have detailed our observing and
  calibration methodology in Kulkarni et al. (1999) and
  Frail et al. (2000).

  Observations at IRAM were made using the Max-Planck Millimeter
  Bolometer (MAMBO; Kreysa et al. 1999) at the IRAM 30-m telescope on
  Pico Veleta, Spain. Observations were made in standard on-off mode.
Gain calibration was performed using observations of Mars, Uranus, and
Ceres. We estimate the calibration to be accurate to 15\%.  Using the
MOPSI software package (Zylka 1998), the temporally correlated
variation of the sky signal (sky-noise)
was subtracted from all bolometer signals.  The source was
observed on March 4, 5, and 9 under very stable atmospheric
conditions, and on March 6 with high atmospheric opacity. From March
24 to 26, the source was briefly re-observed three times for a total
on+off integration time of 2000 sec, but no signal was detected.

\section{The Data}
\label{sec:data}

In Figure~\ref{fig:spectrum} we present broad-band spectra from March
5.66 UT ($\Delta t \approx 4.25$ days) and March 13.58 UT ($\Delta t
\approx 12.17$ days). Radio light curves at 4.86, 8.46, 22.5, and 250
GHz from Table~\ref{tab:Table-VLA} are presented in
Figure~\ref{fig:rlc1}, while optical/IR curves are
shown in Figure~\ref{fig:olc}.

The quoted uncertainties in the flux densities given in
Table~\ref{tab:Table-VLA} report only measurement error and do not
contain an estimate of the effects of interstellar scattering, which
is known to be significant for radio afterglows (e.g. \cite{fks+99}).
We can get some guidance on the expected magnitude of the ISS-induced
modulation of our flux density measurements (in time and frequency)
using the models developed by Taylor and Cordes (1993), Walker (1998), and 
Goodman (1997).  

From the Galactic coordinates of \grb\, ({\it l,b})=($48.7^\circ,44.3^\circ$),
we find, using the Taylor and Cordes model, that the scattering measure is
$SM_{-3.5}\approx 0.7$.  The distance to the scattering screen, $d_{scr}$, 
is one half the distance through the ionized gas layer, 
$d_{scr}=(hz/2)({\rm sin}b)^{-1}\approx 0.72$ kpc, using $hz\approx 1$ kpc. 
From Walker's analysis, the 
transition frequency between weak and strong scintillation is then given by
$\nu_0=5.9 SM_{-3.5}^{6/17} d_{scr}^{5/17} \approx 4.7$ GHz.  Goodman (1997)
uses the same scalings, but with a different normalization for the transition 
frequency, giving a larger value, $\nu_0 \approx 8.3$ GHz.  In this section
we follow Walker's analysis, and note that the numbers from Goodman will give
different results. 

For frequencies larger than the transition frequency the modulation index 
(i.e. the r.m.s. fractional flux variation) is 
$m_{\nu}=(\nu_0/\nu)^{17/12}$, and the modulation timescale in hours is 
$t_\nu \approx 6.7(d_{scr}/\nu)^{1/2}$.  From this analysis we find that 
the modulation index is of order 0.4 at 8.46 GHz, 0.2 at 15 GHz, 0.10 at 
22.5 GHz, and is 
negligible at higher frequencies.  The modulation timescales are of order
2.0 hours at 8.46 GHz, 1.5 hours at 15 GHz, and 1.2 hours at 22.5 GHz.    
It is important to note that factor 2 uncertainties in the scattering 
measure allow the modulation index to vary by $\sim 50\%$.   

At these frequencies the expansion of the fireball will begin to ``quench'' 
the ISS when the angular size of the fireball exceeds the angular size 
of the first Fresnel zone, $\theta_F = 8(d_{scr}\nu_{GHz})^{-1/2}\mu$as.
To describe the evolution of the source size with time, we have used an 
expanding jet model (see \cite{fks+99}),
with the factor $(E_{52}/n_1)^{1/8}$ assumed to be of order unity, which 
gives $\theta_s\approx 3.1(\Delta t_d/15)^{1/2} \mu{\rm as}$.  
Once the source size exceeds the Fresnel size (after approximately two
weeks at 8.46 GHz), the modulation index has to be
corrected by a factor $(\Delta t_d/15)^{-7/12}$.

The measurements at 4.86 GHz occur near the transition frequency and we 
therefore expect $m_{4.86}$ to be large $\sim 0.65-1.0$.  At 1.43 GHz 
the observations were made in the strong regime of ISS where we expect both 
refractive and diffractive scintillation.  Point source refractive scintillation
at 1.43 GHz has a modulation index $m_{1.43,r}=(\nu/\nu_0)^{17/30} \approx 0.5$,
with a timescale of $t_{1.43,r} \approx 2(\nu_0/\nu)^{11/5} \approx 1$ day.  The 
refractive ISS is ``quenched'' when the angular size of the source is larger than 
$\theta_r=\theta_{F0}(\nu_0/\nu)^{11/5}$, where $\theta_{F0}$ is the angular
size of the first Fresnel zone at $\nu_0=4.7$ GHz.  As with weak scattering, 
the modulation index has to be corrected by a factor $(\Delta t_d/15)^{-7/12}$
after this point. The diffractive 
scintillation has a modulation index $m_{1.43,d}=1$, and a timescale,
$t_{1.43,d}\approx 2(\nu/\nu_0)^{6/5}\approx 0.5$ hrs $\ll t_{1.43,r}$.
The source can no longer be approximated by a point source when its angular 
size exceeds
$\theta_d=\theta_{F0}(\nu/\nu_0)^{6/5}$, and correspondingly the modulation 
index has to be corrected by a factor $(\Delta t_d/15)^{-1/2}$.

The redshift of \grb\ was measured using the Hubble Space Telescope 
to be 1.95$\pm$0.1 by Smette et al. (2000) and was later refined by Castro 
et al. (2000) using the Keck II 10-m Telescope to a value of 2.0335$\pm$0.0003.
The combined fluence measured by the GRB detector on board the Ulysses satellite, 
and the X-ray/gamma-ray Spectrometer (XGRS) on board the Near Earth Asteroid 
Rendezvous (NEAR) satellites, in the 25-100 keV
and $>$100 keV bands was $\sim$4$\times 10^{-6}$ erg/cm$^2$.  
Using the cosmological parameters $\Omega_0$=0.3, $\Lambda_0$=0.7 and 
$H_0$=65 km/sec/Mpc,
we find that the isotropic $\gamma$-ray energy release from the GRB was
$E_{iso}\approx 5.4\times 10^{52}$ ergs.

\section{A Self-Consistent Jet Interpretation}
\label{sec:model}

According to the standard, spherical GRB model, the optical light
curves should obey a simple power-law decay, $F_\nu \propto
t^{-\alpha}$, with $\alpha$ changing at most by 1/4 as the electrons
age and cool (Sari, Piran, \& Narayan 1998).  From Figure~\ref{fig:olc} 
it is evident that the optical light
curves steepen substantially ($\Delta\alpha>1/4$) between days 7 and 8, which 
indicates that this burst cannot be described within this standard model of an
expanding spherical blast wave.  This break can be attributed to a jet-like 
or collimated ejecta (Rhoads 1999; Sari, Piran, \& Halpern 1999).

The jet model of GRBs predicts the time evolution of flux from the
afterglow, and of the parameters $\nu_a \propto t^{-1/5}$, $\nu_m
\propto t^{-2}$, and $F_{\nu,max} \propto t^{-1}$.  This model holds
for $t>t_{jet}$, where $t_{jet}$ is defined by the condition
$\gamma(t_{jet}) \sim \theta_0^{-1}$.  Prior to $t_{jet}$ the time
evolution of the afterglow is described by a spherically expanding
blast wave, with the scalings $\nu_a \propto$ const., $\nu_m \propto
t^{-3/2}$, and $F_{\nu,max} \propto$ const.  In this paper we
designate this model as ISM+jet.  Throughout the analysis
we assume that the cooling frequency, $\nu_c$, lies above the optical
band for the entire time period under discussion in this paper.  

\begin{figure*} 
  \centerline{\hbox{\psfig{figure=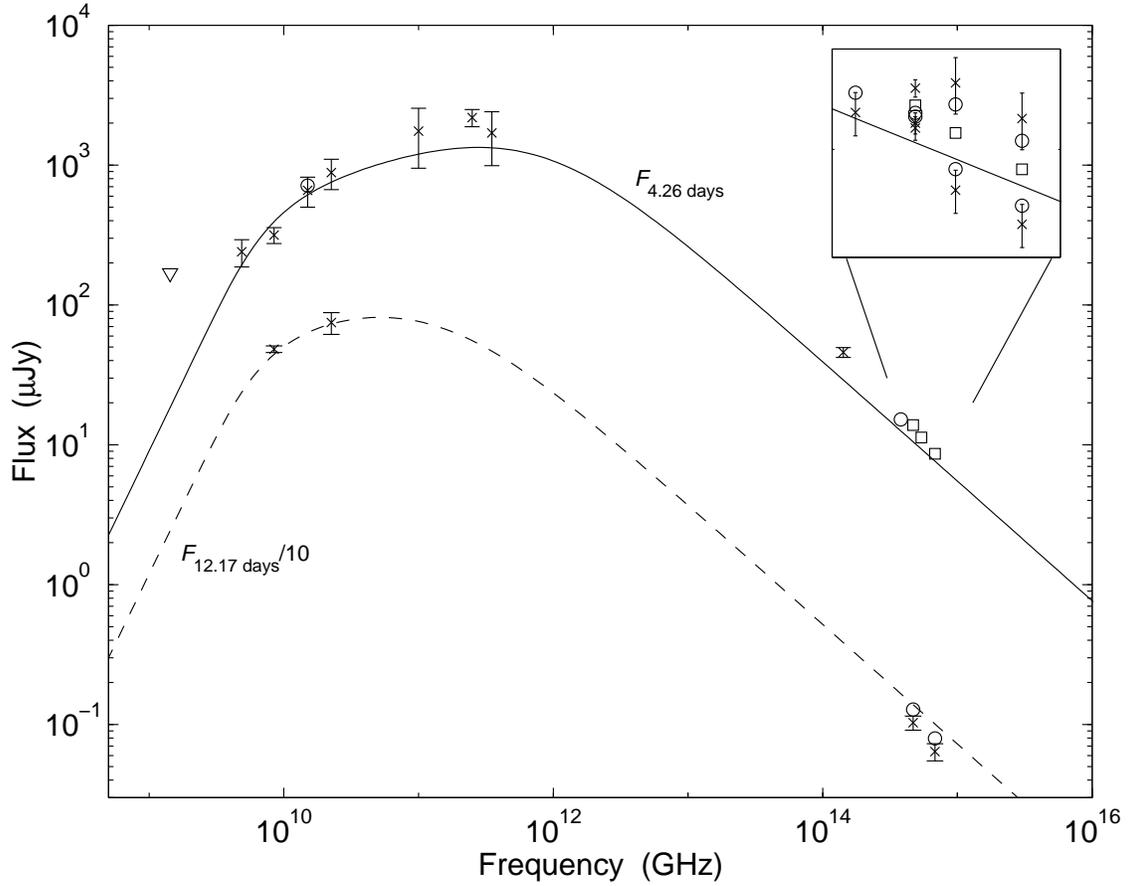,width=15cm}}}
\caption[]{The radio to optical spectral flux distribution of \grb\ on 
  2000 March 5.66 UT ($\Delta t \approx 4.26$ days after the burst),
  and 2000 March 13.58 UT ($\Delta t \approx 12.17$ days). The solid
  and dashed lines are the global fits
  based on the smoothed synchrotron emission spectrum of Granot,
  Piran, and Sari (1999a,b).  The radio measurements are from
  Table~\ref{tab:Table-VLA}.  The optical/IR data are from Rhoads
  and Fruchter (2000), Sagar et al. (2000), and Masetti et al.~(2000), 
  converted to Jansky flux units (\cite{bb88}, \cite{FSI95}),
  and corrected for Galactic foreground
  extinction (\cite{sfd98}), giving $E(B-V)=0.053$.  All data were taken 
  within 0.5 days of the fiducial
  dates, and the circles are the corrections to the data to the fiducial
  times, $\Delta t=4.26$ days, and $\Delta t=12.17$ days.  The squares 
  in the optical bands are weighted averages of multiple measurements
  within 1 day of $\Delta t=4.26$ days (see insert).  The inverted triangle
  is an upper limit at 1.43 GHz from day 4.26.  The data
  points at 100, 250, and 350 GHz are weighted averages of the
  individual measurements from around day 4 (see Table~\ref{tab:Table-VLA}).  
  Note that the data and fit from
  $\Delta t=12.17$ days were divided by a factor of 10 to avoid
  overlap with the $\Delta t=4.26$ curve.
\label{fig:spectrum}}
\end{figure*}

\begin{figure*} 
  \centerline{\hbox{\psfig{figure=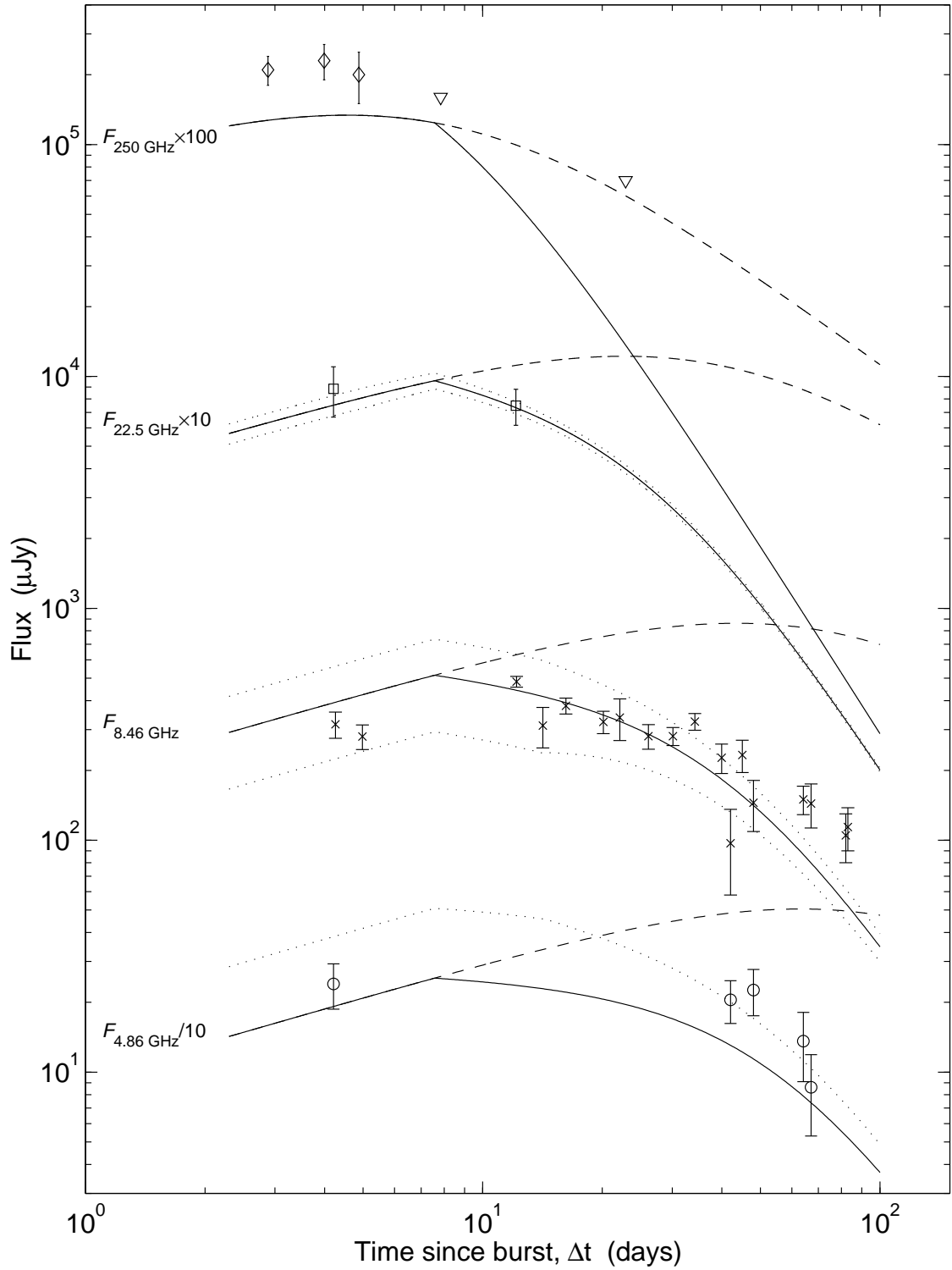,width=15cm}}}
\caption[]{Radio light curves for 4.86, 8.46, 22.5, and 250 GHz.  
  The data were obtained at the VLA and IRAM (see \S{2}).  The
  solid lines are the global fit, based
  on the time dependences of the parameters $\nu_a$, $\nu_m$, and
  $F_{\nu,max}$, and the smoothed, broken power law synchrotron spectrum 
  calculated by Granot, Piran and Sari (see \S{4}).  
  The sharp break at time $t=t_{jet}\approx 7.5$ days corresponds to
  the transition from spherical to jet geometry.  The dashed curve
  shows the prediction for a spherical evolution of the afterglow.  
  The dotted lines indicate the 
  maximum and minimum range of flux expected from ISS (see \S{3}).  
  Note that the data and fit for
  4.86 GHz were divided by a factor of 10, the data and fit for
  22.5 GHz were multiplied by a factor of 10, and the data and fit for
  250 GHz were multiplied by a factor of 100 to avoid overlap between the
  four curves.
\label{fig:rlc1}}
\end{figure*}

\begin{figure*} 
  \centerline{\hbox{\psfig{figure=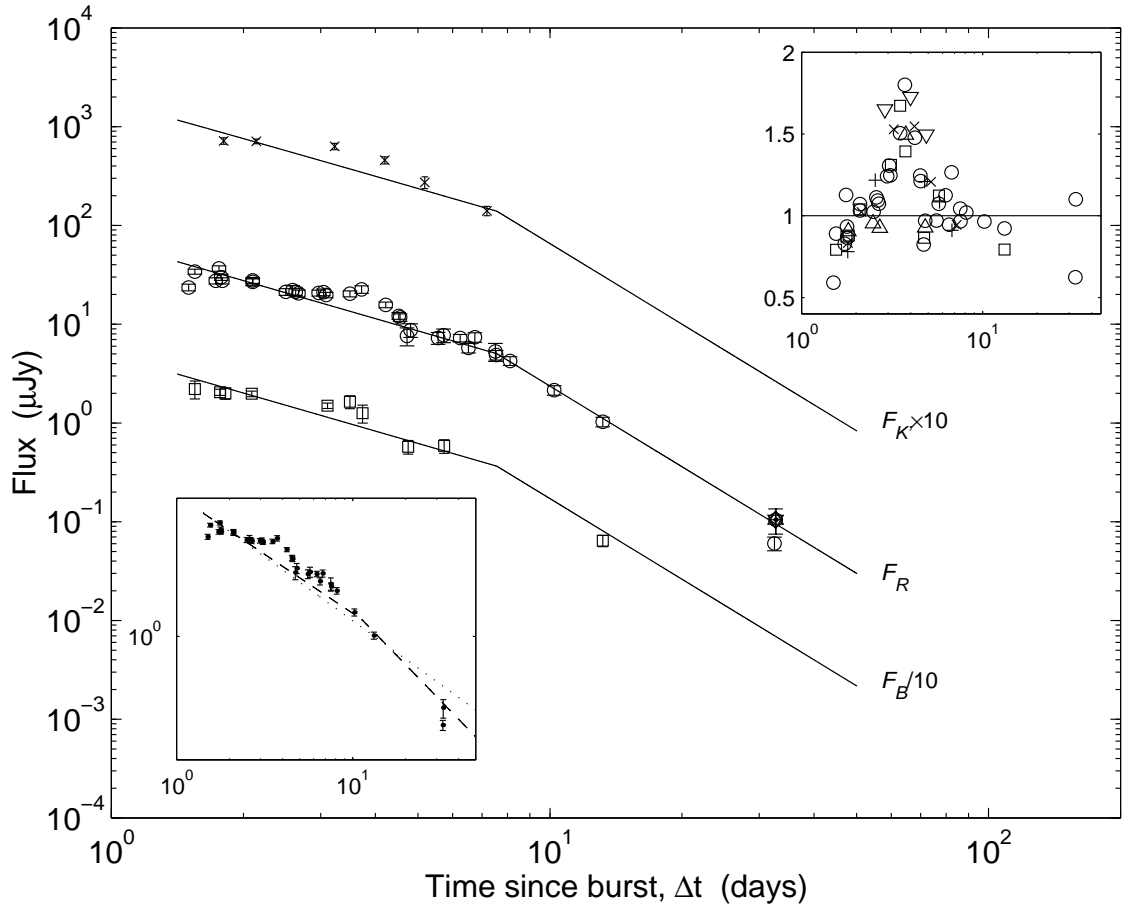,width=15cm}}} 
\caption[]{Optical light curves for B, R, and K' bands of \grb{}.  The
  data are from Rhoads and Fruchter (2000), Sagar et al. (2000),
  Masetti et al.~(2000), and Diercks et al. (2000; filled circle), 
  and contain a correction for Galactic
  extinction (see Figure \ref{fig:spectrum}). Following Masseti et al.
  (2000), we added a 5\% systematic uncertainty in quadrature to all
  optical measurements to account for discrepancies between the
  different telescopes and instruments used in the observations.  The
  solid lines are the global fit, based
  on the smoothed, broken power law synchrotron spectrum calculated by
  Granot, Piran and Sari (see \S{4}).  In the top right insert 
  are plotted the data points divided by the respective model fit for 
  all bands (circles, squares, crosses, triangles, pluses, and inverted 
  triangles indicate R, B, K', V, I, and 250 GHz bands, respectively).  
  It is evident that the short-timescale fluctuations are
  achromatic and with a comparable amplitude in all bands, spanning from optical 
  to radio.  The insert on the bottom left portion of the figure shows the global
  fits based on the wind-only (dotted line), and wind+jet (dashed line) models
  overlaid on the R band data.  It is clear that the steeper decline predicted 
  for a fireball expanding into a wind-shaped circumburst medium results 
  in a much poorer fit relative to the ISM+jet model.  Note that the data
  and fit for B  band were divided by a factor of 10, and that the data 
  and fit for K' band were multiplied by a factor of 10 to avoid overlap
  between the three curves.
\label{fig:olc}}
\end{figure*}

At any point in time the spectrum is roughly given by the broken
power law $F_{\nu} \propto \nu^2$ for $\nu<\nu_a$, $F_{\nu} \propto \nu^{1/3}$ for
$\nu_a<\nu<\nu_m$, and $F_{\nu} \propto \nu^{-(p-1)/2}$ for $\nu>\nu_m$,
where $p$ is the electron power law index.  To globally fit the entire
radio and optical/IR data set we employed the smoothed form of the broken power 
law synchrotron
spectrum, calculated by Granot, Piran, and Sari (1999a,b). With this
approach we treat $t_{jet}$, $p$, and the values of $\nu_a$, $\nu_m$,
and $F_{\nu,max}$ at $t=t_{jet}$ as free parameters.
This method forces $t_{jet}$ to have the same value at all frequencies. 
We find the following values for the burst parameters: $t_{jet}=7.5
\pm 0.5$ days, $p=2.70 \pm 0.04$, $\nu_a(t=t_{jet})=8.3 \pm 1.8$ GHz,
$\nu_m(t=t_{jet})=(3.0 \pm 0.3) \times 10^{11}$ Hz,
$F_{\nu,max}(t=t_{jet})=2.7 \pm 0.2$ mJy, where the errors are
derived from the diagonal elements of the correlation matrix.  We 
note that there is substantial covariance between some of the parameters
and therefore these error estimates should be treated with caution.
From our fit the asymptotic temporal decay slopes of the optical light 
curves are $p$ we find $\alpha_1=-3(p-1)/4=-1.28$ for $t<t_{jet}$, and
$\alpha_2=-p=-2.70$ for $t>t_{jet}$.  The fits are shown in 
figures~\ref{fig:spectrum},~\ref{fig:rlc1},~and \ref{fig:olc}.

The total value of $\chi^2$ for the global fit is poor. We obtain
$\chi^2=450$ for 94 degrees of freedom. The bulk of this value, 340,
comes from the 61 optical data points, and is the result of the 
fluctuations, which are not accounted for by our model. The radio data 
contribute a
value of 110 to $\chi^2$ for $38$ data points. This is probably the
result of scintillation. If we increase the
errors to accommodate for the expected level of scintillation (see \S{3}) 
we obtain a good fit with $\chi_{radio}^2=39/33$ degrees of freedom.
 
From figure~\ref{fig:spectrum} it is clear that the global fit accurately 
describes the broad-band spectra from days 4.26 ($<t_{jet}$)
and 12.17 ($>t_{jet}$), with a single value of $p=2.70$, which rules out 
the possibility that the steepening of the light curves at $t=t_{jet}$ 
is the result of a time-varying $p$.  

Trying to model the data using the approach outlined above, but for a 
wind-shaped 
circumburst medium results in a poor description of the data, because  
the wind model does not exhibit a break, although 
one is clearly seen in the optical data.  As a result, the model fit is too low at
early times, and too high at late times relative to the data (see insert in
figure~\ref{fig:olc}).  The value of $\chi^2$ for the wind model relative to 
 the ISM+jet model described above is $\chi^2_{wind}/\chi^2_{ISM+jet} \sim 3$.
Therefore, a wind-shaped model can be ruled out as a description of 
the afterglow of \grb{}.

A jet evolution combined with a wind-shaped circumburst medium provides a 
more reasonable fit 
than a wind only model.  The wind evolution of the fireball will only
be manifested for $t<t_{jet}$ since once $\gamma(t_{jet})\approx\theta_0^{-1}$
the jet will expand sideways and appear to observers as if it were expanding
into a constant density medium (Chevalier \& Li 1999b; Livio \& Waxman 1999).
The resulting parameters from such a fit differ considerably from
the parameters for the ISM+jet model quoted above, and the relative
value of $\chi^2$ between the two models is 
$\chi^2_{wind+jet}/\chi^2_{ISM+jet} \sim 2$.  This model suffers from a serious 
drawback in its
description of the optical/IR light curves.  Because the predicted decay of  
these light curves prior to $t_{jet}$ is steeper than in the ISM+jet model, 
the model fit, from 2 days after the burst up to the break time, is too low 
relative to the data (see insert in figure~\ref{fig:olc}).

The approach outlined above, which uses one power law temporal
evolution of $\nu_a$, $\nu_m$, and $F_{\nu,max}$ before $t_{jet}$, and
a different power law evolution after $t_{jet}$, creates a sharp break
at $t_{jet}$ as seen in
Figures~\ref{fig:rlc1}, and~\ref{fig:olc}. In contrast,
a smooth analytical form for the jet break, was used by several other
groups (Harrison et al. 1999; Stanek et al. 1999; Israel et al.
1999; Kuulkers et al. 2000) to describe the afterglow of
GRB\thinspace{990510}. Beuermann et al. (1999), suggested that since
there is no detailed theory for the jet transition, the shape of the break
(i.e. how smooth or sharp it is) should be kept as a free parameter. 
In a recent paper, Rhoads and Fruchter (2000) used the same approach, 
a free shape-parameter, and
got a smooth light curve as their best fit. However, they have not
forced the relation between the asymptotic temporal decay slopes
before and after $t_{jet}$, ($\alpha_1$ and $\alpha_2$, respectively) 
and they allowed separate slopes and break
times for the R and K' bands. Using the Beuermann et al. formula, forcing
the asymptotic relations between $\alpha_1$ and $\alpha_2$ to fit the
theory, and using a single $t_{jet}$ for both bands, we find that the
best fit is a sharp break. This may be the result of the unexplained
fluctuations that appear in the optical bands. In the radio regime,
there is no data around $t_{jet}$ and therefore a smooth connection
would not make a difference. This justifies the use of a sharp break
in our fit.

The global fitting approach has several advantages over fitting each
component of the data set independently.  For example, the K' data is
only available up to day 7.18 ($\lesssim t_{jet}$) after the burst.
Therefore, by fitting it independently of the R band and of the radio data
we cannot find $t_{jet}$, if it is indeed after $7$ days. Moreover,
since, as Masetti et al., and Sagar et al. claim, and as we can see
from Figure~\ref{fig:olc}, there is an additional process which
superposes achromatic fluctuations with an overall rise and decline
centered on day 3, on top of the smoothly
decaying optical emission (see insert in figure~\ref{fig:olc}), 
then fitting the K' data independently will
confuse this behavior with the jet break.  This explains the
result of Rhoads and Fruchter of $t_{jet,K'}\sim3$ days. 
It is worth noting that fitting the available R band data from before day 8 by
itself, gives a value of $t_{jet,R}\sim 3.5 \, {\rm days} \sim t_{jet,K'}$. 

Simultaneous fitting of the entire data set makes it
possible to study the overall behavior of the fireball regardless of
any additional sources of fluctuations, because the large range in
frequency and time of the data reduces the influence of such
fluctuations. Remarkably, using this global fit with only the radio data, 
ignoring the optical observations, we obtain 
$t_{jet,Radio}\approx 7.7\, {\rm days} \sim t_{jet}$.  Thus, the radio data
serves to support the jet model, and provides an additional estimate
for the jet break time, independent of the somewhat ambiguous optical
data.

From the global fit we find the first self-consistent indication that the short-timescale
optical fluctuations are achromatic, even in the K' band (see insert in 
figure~\ref{fig:olc}).  By simply dividing the B, R, V, I, and K' data by the values
from the global fit we find that the fluctuations happen simultaneously and 
with similar amplitudes in all bands.  Moreover, the overall structure of the
fluctuations is a sharp rise and decline centered on day 4, and with an 
overall width of 3.5 days, which gives $\delta t/t \sim 1$, where $\delta t$
is the width of the bump.  The optical/IR data starts at day 1.5 lower by
25-50\% than the model fit, then rises to a peak level of 50-75\% relative to
the model at day 4, and drops to the predicted level at about day 5, at which point
it follows the predicted decline of the ISM+jet model.  
  
It is interesting to note that the 250 GHz data, which is not affected by 
ISS-induced fluctuations, also shows a peak amplitude approximately 70\% higer 
than the model fit around day 4 (see insert in figure~\ref{fig:olc}).  
At the lower radio frequencies there are not enough data points 
to discern a similar behavior.  Moreover, at these frequencies it 
would have been difficult to disentangle such fluctuations from ISS-induced
fluctuations in any case.  The large range
in frequency of this achromatic fluctuation, coupled with the similar
level of absolute deviation from the model fit suggests that it is the
result of a real physical process.

It is possible to explain this fluctuation as a result of a non-uniform
ambient density.  The value of $\nu_m$ is independent of the ambient medium 
density, and since $F_{\nu,max} \propto n_1^{1/2}$, we expect the flux at
frequencies larger than $\nu_a$ to vary achromatically, and with the same
amplitude, $F_{\nu} \propto n_1^{1/2}$.  For frequencies lower than $\nu_a$ we have 
to take into account the density dependence $\nu_a \propto n_1^{3/5}$ so that
the flux will vary according to 
$F_{\nu} \propto F_{\nu,max}\nu_a^{-2} \propto n_1^{-7/10}$.  This means that for
frequencies lower than $\sim 11$ GHz we actually expect the flux to 
fluctuate downward at the same time that it fluctuates upward at higher
frequencies.  In practice, we don not have enough data around this time to confirm this
behavior.  In order to match the observed amplitude of the fluctuation $\sim 80$\%, 
the ambient density has to vary by about a factor of 3. 

Using the value of $t_{jet}$ from our global fit, we can calculate
the jet opening angle, $\theta_0$, from the equation:
\begin{equation}
\theta_0 \approx 0.05(t_j/hr)^{3/8}(1+z)^{-3/8}(n_1/E_{52})^{1/8},
\label{eqn:open}
\end{equation}
(Sari et al. 1999; Livio \& Waxman 1999) where $E_{52}$ is the isotropic energy release,
which can be roughly estimated from the observed fluence (using the equations from 
Rhoads (1999) results in a smaller opening angle). 
 From this equation, we calculate a value
of $\theta_0 \approx 0.2n_1^{1/8}$ radians. This means that the actual
energy release from \grb\ is reduced by a factor of 50 relative
to the isotropic value, $E_{iso}\approx 5.4\times 10^{52}$ ergs,  
which gives $E=1.1\times 10^{51}n_1^{1/4}$ ergs.

\section{Conclusion}
\label{sec:conc}

The afterglow emission from \grb\ can be well described in the
framework of the jet model of GRBs.  Global fitting of the radio and
optical data, allows us to calculate the values of $p$, $t_{jet}$, and
the time evolution of $\nu_a$, $\nu_m$, and $F_{\nu,max}$ in a
self-consistent manner. Within this approach the proposed discrepancy
between the behaviors of the R band and K' band light curves,
suggested by Rhoads and Fruchter, is explained as the result of the
lack of data for $t>7.18$ days $(\lesssim t_{jet})$ at K', and the
existence of achromatic substructure from fluctuations in the radio and
optical/IR regimes. The value for the break time from the global,
self-consistent approach we have used is $t_{jet}=7.5$ days at all
frequencies.

The long-lived radio emission from the burst, spanning a large range
in frequency and time, plays a significant role in our ability to
extract the time evolution of $\nu_a$, $\nu_m$, and $F_{\nu,max}$ from
the data. In the case of this GRB in particular, the large range in
frequency and time is crucial, since it serves to reduce the effects
of unexplained deviations from the simple theory, such as the
short-timescale fluctuations in the optical bands, on the overall
evolution of the fireball.

We end with some words of caution. In our analysis we assumed that
$\nu_c$ lies above the optical band throughout the evolution of the
fireball, and successfully got a reasonable fit to the data. However, it is 
possible that
another set of parameters, with $\nu_c$ below the optical band, can
fit the data equally well. Preliminary work in this direction
indicates that the gross features of the fireball evolution (e.g. a
break time of $\sim$7 days) remain unaltered.

\acknowledgements Research at the Owens Valley Radio Observatory is
supported by the National Science Foundation through NSF grant number
AST 96-13717.

\begin{deluxetable}{lrccc}
\tabcolsep0in\footnotesize
\tablewidth{\hsize}
\tablecaption{Radio Observations of \grb\ \label{tab:Table-VLA}}
\tablehead {
\colhead {Epoch}      &
\colhead {$\Delta t$} &
\colhead {Telescope} &
\colhead {$\nu_0$} &
\colhead {S$\pm\sigma$} \\
\colhead {(UT)}      &
\colhead {(days)} &
\colhead {} &
\colhead {(GHz)} &
\colhead {($\mu$Jy)}
}
\startdata
2000 March $\phantom{0}$4.29 & 2.88   & IRAM & 250 & 2100$\pm$300 \nl
2000 March $\phantom{0}$4.75 & 3.34   & JCMT & 350 & 3736$\pm$3700 \nl
2000 March $\phantom{0}$4.98 & 3.57   & Ryle & 15.0 & 660$\pm$160 \nl
2000 March $\phantom{0}$5.41 & 4.00   & IRAM & 250 & 2300$\pm$400 \nl
2000 March $\phantom{0}$5.53 & 4.12   & JCMT & 350 & 2660$\pm$1480 \nl
2000 March $\phantom{0}$5.57 & 4.16   & OVRO & 100 & 2850$\pm$950  \nl
2000 March $\phantom{0}$5.67 & 4.26   & VLA & 1.43 & 11$\pm$79    \nl
2000 March $\phantom{0}$5.67 & 4.26   & VLA & 4.86 & 240$\pm$53  \nl
2000 March $\phantom{0}$5.67 & 4.26   & VLA & 8.46 & 316$\pm$41  \nl
2000 March $\phantom{0}$5.67 & 4.26   & VLA & 22.5 & 884$\pm$216  \nl
2000 March $\phantom{0}$6.29 & 4.88   & IRAM & 250 & 2000$\pm$500 \nl
2000 March $\phantom{0}$6.39 & 4.98   & VLA & 8.46 & 289$\pm$34  \nl
2000 March $\phantom{0}$6.50 & 5.09   & JCMT & 350 & 1483$\pm$1043 \nl
2000 March $\phantom{0}$6.57 & 5.16   & OVRO & 100 & $-$99$\pm$1500  \nl
2000 March $\phantom{0}$9.25 & 7.84   & IRAM & 250 & 400$\pm$600 \nl
2000 March $\phantom{0}$10.21 & 8.80  & Ryle & 15.0 & 480$\pm$300 \nl
2000 March $\phantom{0}$13.58 & 12.17 & VLA & 8.46 & 483$\pm$26   \nl
2000 March $\phantom{0}$13.58 & 12.17 & VLA & 22.5 & 748$\pm$132  \nl
2000 March $\phantom{0}$15.58 & 14.17 & VLA & 8.46 & 312$\pm$62  \nl
2000 March $\phantom{0}$17.61 & 16.20 & VLA & 8.46 & 380$\pm$29 \nl
2000 March $\phantom{0}$21.52 & 20.12 & VLA & 8.46 & 324$\pm$36 \nl
2000 March $\phantom{0}$23.55 & 22.14 & VLA & 8.46 & 338$\pm$69   \nl
2000 March $\phantom{0}$24.29 & 22.88 & IRAM & 250 & $-$300$\pm$500 \nl
2000 March $\phantom{0}$27.55 & 26.14 & VLA & 8.46 & 281$\pm$34   \nl
2000 March $\phantom{0}$31.53 & 30.12 & VLA & 8.46 & 281$\pm$25  \nl
2000 April $\phantom{0}$4.59 & 34.18 & VLA & 8.46 & 325$\pm$27   \nl
2000 April $\phantom{0}$10.36 & 39.95 & VLA & 8.46 & 227$\pm$33  \nl
2000 April $\phantom{0}$12.47 & 42.06 & VLA & 4.86 & 210$\pm$43   \nl
2000 April $\phantom{0}$12.47 & 42.06 & VLA & 8.46 & 91$\pm$38   \nl
2000 April $\phantom{0}$15.43 & 45.02 & VLA & 8.46 & 233$\pm$37  \nl
2000 April $\phantom{0}$18.47 & 48.06 & VLA & 4.86 & 226$\pm$51  \nl
2000 April $\phantom{0}$18.47 & 48.06 & VLA & 8.46 & 145$\pm$36  \nl
2000 May $\phantom{0}$4.49 & 64.13 & VLA & 4.86 & 136$\pm$45  \nl
2000 May $\phantom{0}$4.49 & 64.13 & VLA & 8.46 & 150$\pm$20  \nl
2000 May $\phantom{0}$7.50 & 67.09 & VLA & 4.86 & 85$\pm$33  \nl
2000 May $\phantom{0}$7.50 & 67.09 & VLA & 8.46 & 144$\pm$31  \nl
2000 May $\phantom{0}$22.45 & 82.04 & VLA & 8.46 & 105$\pm$25  \nl
2000 May $\phantom{0}$23.45 & 83.04 & VLA & 8.46 & 114$\pm$24  \nl
\enddata
\tablecomments{The columns are (left to right), (1) UT date of the
  start of each observation, (2) time elapsed since the $\gamma$-ray
  burst, (3) telescope name, (4) observing frequency, and (5) peak
  flux density at the best fit position of the radio transient, with
  the error given as the root mean square noise on the image.  The
  JCMT observations did not detect the source at each epoch
  individually, but by averaging the 3.875 hours of integration over
  the three epochs, we obtain a 2.5$\sigma$ detection of 1.70 $\pm$
  0.71 mJy.}
\end{deluxetable}

\clearpage 

\clearpage

\clearpage 

\end{document}